\documentclass[twocolumn]{aastex62}
\usepackage{comment}
\usepackage{amsmath}	
\usepackage{amssymb}
\usepackage{footnote}
\usepackage{appendix}
\usepackage[graphicx]{realboxes}
\usepackage{threeparttable}
\usepackage[caption=false]{subfig}

\newcommand{\oii}{[O~{\sc ii}]}
\newcommand{\oiii}{[O~{\sc iii}]}

\newcommand{\nii}{[N~{\sc ii}]}

\graphicspath{{./}{figures/}}

\received{October 08, 2019}
\revised{December 11, 2019}
\accepted{December 23, 2019}
\submitjournal{ApJL}
\shorttitle{ASKAP FRB host galaxies}
\shortauthors{Bhandari et al.}

\begin{document}

\title{The host galaxies and progenitors of Fast Radio Bursts localized with the Australian Square Kilometre Array Pathfinder 
}

\correspondingauthor{Shivani Bhandari}
\email{shivani.bhandari@csiro.au}
\author[0000-0003-3460-506X]{Shivani Bhandari}
\affil{Australia Telescope National Facility, CSIRO Astronomy and Space Science, PO Box 76, Epping, NSW 1710, Australia }

\author[0000-0002-1136-2555]{Elaine M. Sadler}
\affil{Australia Telescope National Facility, CSIRO Astronomy and Space Science, PO Box 76, Epping, NSW 1710, Australia }
\affiliation{Sydney Institute for Astronomy, School of Physics A28, The University of Sydney, NSW 2006, Australia}

\author{J. Xavier Prochaska}
\affil{University of California, Santa Cruz, 1156 High St., Santa Cruz, CA 95064, USA}
\affiliation{
Kavli Institute for the Physics and Mathematics of the Universe (Kavli IPMU),
5-1-5 Kashiwanoha, Kashiwa, 277-8583, Japan}

\author[0000-0003-3801-1496]{Sunil Simha}
\affil{University of California, Santa Cruz, 1156 High St., Santa Cruz, CA 95064, USA}

\author[0000-0003-4501-8100]{Stuart D. Ryder}
\affiliation{Department of Physics and Astronomy, Macquarie University, NSW 2109, Australia }

\author[0000-0003-1483-0147]{Lachlan Marnoch}
\affil{Department of Physics and Astronomy, Macquarie University, NSW 2109, Australia }
\affiliation{Australia Telescope National Facility, CSIRO Astronomy and Space Science, PO Box 76, Epping, NSW 1710, Australia }

\author[0000-0003-2149-0363]{Keith W. Bannister}
\affil{Australia Telescope National Facility, CSIRO Astronomy and Space Science, PO Box 76, Epping, NSW 1710, Australia }

\author[0000-0001-6763-8234]{Jean-Pierre Macquart}
\affiliation{International Centre for Radio Astronomy Research, Curtin University, Bentley WA 6102, Australia }

\author{Chris Flynn}
\affiliation{Centre for Astrophysics and Supercomputing, Swinburne University of Technology, John St, Hawthorn, VIC 3122, Australia}

\author[0000-0002-7285-6348]{Ryan ~M. Shannon}
\affiliation{Centre for Astrophysics and Supercomputing, Swinburne University of Technology, John St, Hawthorn, VIC 3122, Australia}

\author[0000-0002-1883-4252]{Nicolas Tejos}
\affiliation{Instituto de F\'isica, Pontificia Universidad Cat\'olica de Valpara\'iso,
Casilla 4059, Valpara\'iso, Chile}

% Alphabetical order starts now.
\author{Felipe Corro-Guerra}
\affiliation{Instituto de F\'isica, Pontificia Universidad Cat\'olica de Valpara\'iso,
Casilla 4059, Valpara\'iso, Chile}

\author[0000-0002-8101-3027]{Cherie K.Day}
\affiliation{Centre for Astrophysics and Supercomputing, Swinburne University of Technology, John St, Hawthorn, VIC 3122, Australia}
\affiliation{Australia Telescope National Facility, CSIRO Astronomy and Space Science, PO Box 76, Epping, NSW 1710, Australia }

\author{Adam T.Deller}
\affiliation{Centre for Astrophysics and Supercomputing, Swinburne University of Technology, John St, Hawthorn, VIC 3122, Australia}

\author{Ron Ekers}
\affil{Australia Telescope National Facility, CSIRO Astronomy and Space Science, PO Box 76, Epping, NSW 1710, Australia }
\affiliation{International Centre for Radio Astronomy Research, Curtin University, Bentley WA 6102, Australia }

\author{Sebastian Lopez}
\affil{Departamento de Astronom\'ia, Universidad de Chile, Casilla 36-D, Santiago, Chile}

\author{Elizabeth K. Mahony}
\affil{Australia Telescope National Facility, CSIRO Astronomy and Space Science, PO Box 76, Epping, NSW 1710, Australia }

\author{Consuelo Nuñez} 
\affiliation{Instituto de F\'isica, Pontificia Universidad Cat\'olica de Valpara\'iso,
Casilla 4059, Valpara\'iso, Chile}

\author{Chris Phillips}
\affil{Australia Telescope National Facility, CSIRO Astronomy and Space Science, PO Box 76, Epping, NSW 1710, Australia }

\begin{abstract}
The Australian SKA Pathfinder (ASKAP) telescope has started to localize Fast Radio Bursts (FRBs) to arcsecond accuracy from the detection of a single pulse, allowing their host galaxies to be reliably identified. We discuss the global properties of the host galaxies of the first four FRBs localized by ASKAP, which lie in the redshift range $0.11<z<0.48$. All four are massive galaxies (log( $M_{*}/ M_{\odot}$) $\sim 9.4 -10.4$) with modest star-formation rates of up to $2$~$M_{\odot}$~yr$^{-1}$ --- very different to the host galaxy of the first repeating FRB~121102, which is a dwarf galaxy with a high specific star-formation rate. The FRBs localized by ASKAP typically lie in the outskirts of their host galaxies, which appears to rule out FRB progenitor models that invoke active galactic nuclei (AGN) or free-floating cosmic strings. The stellar population seen in these host galaxies also disfavors models in which all FRBs arise from young magnetars produced by superluminous supernovae (SLSNe), as proposed for the progenitor of FRB~121102. A range of other progenitor models (including compact-object mergers and magnetars arising from normal core-collapse supernovae) remain plausible.  

\end{abstract}

\section{Introduction} \label{sec:intro}
Fast Radio Bursts (FRBs) are bright, millisecond-scale duration radio emissions of unknown origin arising at cosmological distances \citep{Cordes2019}. Discovered at the Parkes radio telescope \citep{Lorimer}, the number of facilities around the world that have found FRBs has grown steadily in the last half-decade \citep{Spitler,mls+15,ASKAP,Amiri2019a,Farah2019,Ravi+19}. Most recently, sensitive and wide field-of-view searches for FRBs have come online, and led to a significant increase in the discovery rate \citep{Shannon2018,Amiri2019a}. 

%~~~~~~~~~~~~~~~~~~~~ Figure 1~~~~~~~~~~~~~~~~~~~~~~
\begin{figure*}
\centering
\includegraphics[width=1.1\textwidth]{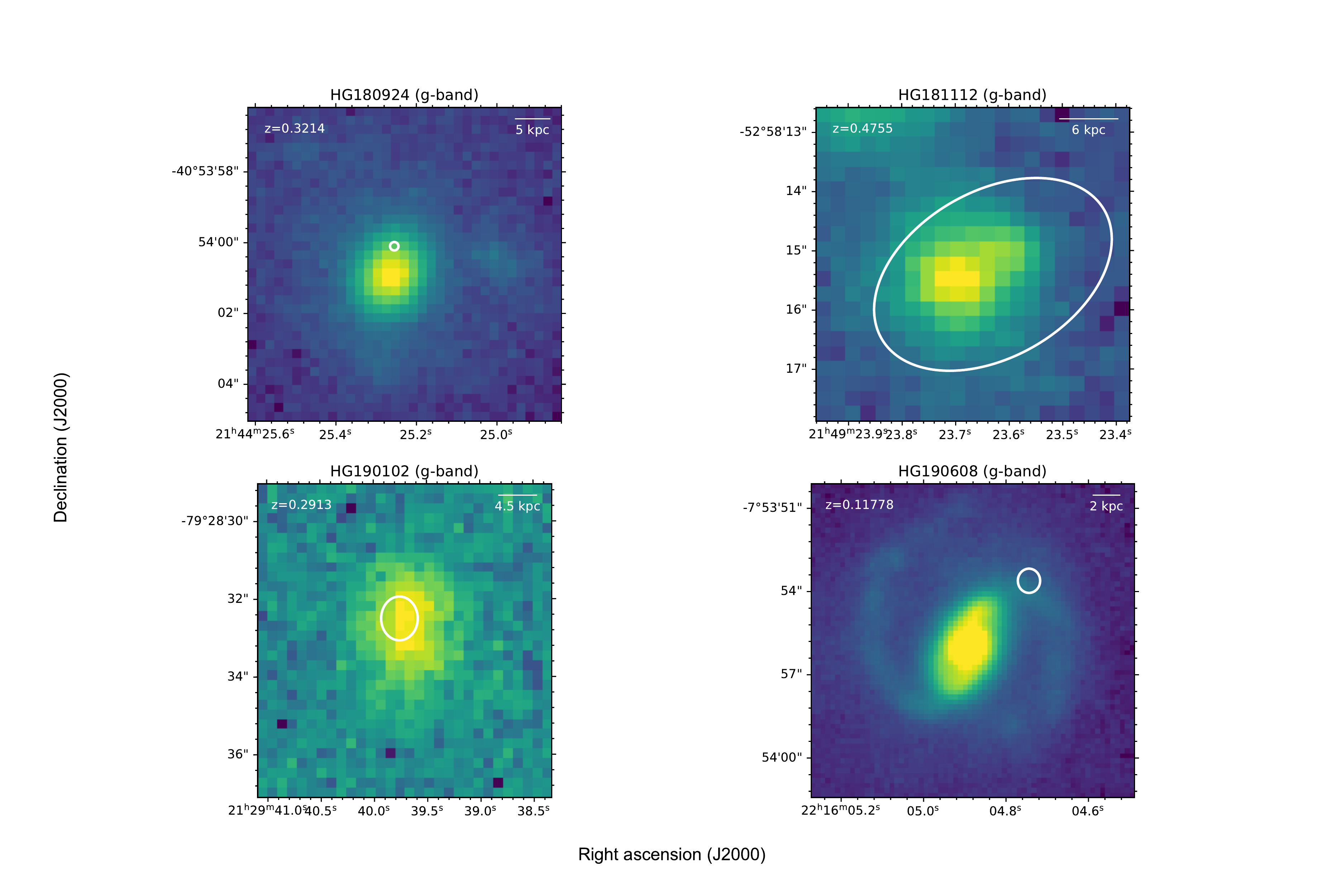}
\caption{g-band FORS2/X-Shooter images of the host galaxies for a sample of localized FRBs (FRB~180924, FRB~181112, FRB~190102 and FRB~190608), overplotted with the positions of each FRB. The white circle/ellipse represents the total uncertainty in the FRB position.  
}
\label{fig:others}
\end{figure*}
%~~~~~~~~~~~~~~~~~~~~~~~~~~~~~~~~~~~~~~~~~~~~~~~~

FRBs show extremely luminous coherent radiation of brightness temperature $T_{\rm b} \sim 10^{35}$K, whose $\sim0.1-10$ ms duration confines their emission regions to $<30-3000$ km. A wide range of theories have been advanced to account for these properties, from those involving supernovae in which the FRB is a feature of a young, expanding supernova remnant \citep{Connor,antony} and super-luminous supernovae \citep{Metzger2017}; the merger/collision of two compact objects such as binary neutron stars (NS-NS) \citep{Yamasaki2018,Totani}, binary white dwarfs (WD-WD) \citep{Kash} and
white dwarf-black hole mergers (WD-BH), the latter via the reconnection of magnetic material \citep{Li2018}; energetic activities from isolated compact objects such as giant pulses from extragalactic pulsars \citep{cordesNS}; giant flares from magnetars \citep{Popov2010,Pen2015}; collision/interaction of neutron stars with AGN \citep{Vieyro2017} and NS ``combing'' \citep{Zhang2017}; collapse of supramassive neutron stars \citep{Falcke2014}; superconducting cosmic strings \citep{Cai+12,CosmicStrings}; and alien beams driving light sails \citep{Aliens}.

The localization of FRBs to host galaxies (HG) will enable exploration of the host galaxy population, their global properties and local FRB environment, which is crucial in understanding FRB progenitor systems. The first localization was achieved for the first known ``repeating FRB" 121102 \citep{spitler2016repeating}. It resides in a low-luminosity ($M_r \approx -17$), star-forming dwarf galaxy at a redshift of $z=0.192$ \citep{VLAlocalisation,host}. Its spatial coincidence with an active star-forming region in the host, a persistent high luminosity radio source and an extreme magnetoionic environment led to a ``concordance picture" of the source of FRB~121102 as a flaring magnetar embedded in a magnetized ion-electron wind nebula \citep{Margalit+2018}.

Despite the recent discovery of the further repeating burst sources by the the Canadian Hydrogen Intensity Mapping Experiment (CHIME) telescope and the Australian Square Kilometre Array Pathfinder (ASKAP) \citep{andersen2019chime,kumar2019faint}, the FRB population is still dominated by one-off and ostensibly non-repeating events. They currently dominate statistical analyses of the FRB phenomenon, and they might have a different progenitor type and arise from cataclysmic implosions or mergers. 

Recently some of these one-off events from ASKAP \citep{Bannister+19,X+19,JP+19} and the Deep Synoptic Array (DSA-10) \citep[FRB~190523]{Ravi+19} have now been localized to Milky Way-like galaxies that are much more massive than the host galaxy of FRB~121102, suggesting that FRBs have diverse host galaxies. 

The Commensal Real-Time ASKAP Fast Transients \citep[CRAFT]{mbb+10} survey is the wide field-of-view FRB search program operating with ASKAP. Initially running in a fly's-eye mode (in which multiple dishes are used, each pointing in different locations on the sky), the project found 25 FRBs\footnote{An additional burst was found in offline incoherent sum search \citep{Agarwal2019} } with a range of dispersion measures (DMs), luminosities and widths in the latter half of 2018 \citep{Shannon2018,Macquart_Spec+19,Bhandari+19,Qiu+19}.  Since late 2018, CRAFT has been operating the facility in incoherent-sum (ICS) mode, where all operating dishes are pointing to the same location and the signals are combined incoherently. FRBs detected in this way trigger a voltage download across all dishes which can be correlated and sub-arcsecond localizations of the FRB positions become possible. 

In this paper, we examine global properties of the host galaxies for the first four ASKAP-localized FRBs, namely FRB~180924, FRB~181112, FRB~190102 and FRB~190608. In Section 2, we describe the optical follow-up observations and the derived host galaxy properties. Section 3 presents their radio follow-up observations and properties. Section 4 discusses the comparison of the properties with other localized FRBs along with FRB models that are ruled out (or favored) by our observations and results. We conclude and provide a summary of our findings in Section 5.

\begin{table*}[!h]
	\centering
	\label{tab:host}
    \resizebox{13.5cm}{!}{
       \rotatebox{90}{
    \begin{threeparttable}
    \begin{tabular}{lcc|cccc}
 \hline
Properties & FRB 121102 & FRB 190523 & FRB 180924 & FRB181112 &	FRB 190102 & FRB 190608 \\
& \citet{host} & \citet{Ravi+19} & \citet{Bannister+19} & \citet{X+19} & \citet{JP+19} & \citet{JP+19} \\
  & & & & & &\\
  \hline
Detection telescope & Arecibo & DSA-10 & ASKAP-ICS & ASKAP-ICS & ASKAP-ICS & ASKAP-ICS \\ 
\textbf{FRB} & & & & & &\\
DM (pc~cm$^{-3}$) & 557(2) & 760.8(6)&	362.4(2) &	589.0(3) &	364.5(3) & 339.5(5)\\
DM$_{\rm ISM}~\rm NE2001 $ (pc~cm$^{-3}$) & 188.0 & 37.0 & 40.5 & 40.2 & 57.3  & 37.2  \\
RA (J2000)& 05h31m58.698(8)s & 13h48m15.6(2)s & 21h44m25.255(8)s & 21h49m23.63(24)s  & 21h29m39.76(17)s & 22h16m4.74(3)s\\
DEC (J2000) & $+$33d08$'$52.6(1)$''$& +72d28$'$11(2)$''$ &$-$40d54$'$00.1(1)$''$ & $-$52d58$'$15.4(14)$''$ & $-$79d28$'$32.5(6)$''$ & $-$07d53$'$53.6(4)$''$\\
RM (rad~m$^{-2}$) & $\sim10^{5}$& - & 14 &10 &110 & -\\
Fluence (Jy\,ms) & $>0.1$ & 280 & 16(1) & 26(3) & 14(1) & 26(4) \\
Pulse width (ms)& 3.0(5) & 0.42(5) & 1.76(9) & 2.1(2) & 1.7(1) & 6.0(8) \\
Energy \tnote{a} (erg~Hz$^{-1}$)
& $3.7\times10^{29}$ & $5.6 \times  10^{33}$ & $3.5 \times 10^{31}$ & $1.0 \times 10^{32}$  & $2.3\times 10^{31}$ & $7.6 \times 10^{30}$ \\     
& & & & & &\\
\textbf{Host galaxy } & & & & & & \\  
Redshift & 0.19273(8) & 0.660(2)& 0.3214(2)& 0.4755(2)& 0.2913(2) & 0.11778 \\
$u-r$ (restframe) \tnote{b}
& - & 1.95 & 1.75(16) & 1.15(15) & 1.33(22) & 1.49(11)\\
M$_{\rm r}$ (restframe) & $-$17.0(2) & $-$22.05(12) & $-$20.76(5) & $-$20.34(7) & $-$19.90(7) & $-$21.15(5) \\
Galactic $E(B-V)$ \tnote{c} 
& - & - & $0.0165$ & $0.0174$ & $0.1994$ & - \\	
log stellar mass ($M_{\odot}$)&	 7.6--7.8 & 11.07(6) & 10.35(14) &	9.42(20)  & 9.50(35) & 10.40(14) \\	
Star formation rate \tnote{d}
($M_{\odot}$~yr$^{-1}$) & 0.4&	$<$ 1.3(2) &$<$ 2&	0.6 & 1.5 & 1.2 \\	log sSFR (yr$^{-1}$) & $-8.14$ & $<-10.95$ &$<-10.0$ & $-9.62$ &$-9.33$ & $-10.30$ \\
Metallicity $Z$ & - & - & 0.008(10) & 0.006(12) & 0.011(16) & 0.009(14) \\
Projected offset from galaxy center (kpc)	& $-$ & $-$ & $3.5(9)$	& $3.1^{+15.7}_{-3.1}$  & $1.5^{+3.4}_{-1.5}$ & $6.8(13)$\\
Half-light radius (kpc) & $-$ & $-$ & 2.5(3) & 8(3) & 5.3(1)& NA\tnote{e} \\
log S$_{\rm radio}$ (6.5\,GHz, W~Hz$^{-1}$) & 22.27 &- & $<21.73$ & $<22.11$  & $<21.62$ & $<20.55 $ \\
F$_{\rm radio}$ (6.5\,GHz, $\mu \rm Jy$) & $\sim 200$ & - & $<21$ & $<21$ & $<19$ & $<10.5$ \\
      \hline
      \caption{Comparison of the properties of the ASKAP FRB host galaxies with the host galaxies of FRB~121102 and FRB~190523. 
	}
      \end{tabular}
      \begin{tablenotes}
\item [a] The energies are derived assuming a flat spectrum for FRBs ($\alpha=0$) and zero k-correction.
\item [b] Restframe colors are derived using CIGALE.
\item [c] Galactic reddening values along host galaxy lines-of-sight estimated using the \citet{SandF} extinction law and the IRSA dust tool.
\item [d] Derived from the dust-corrected H$\alpha$ luminosity using the standard scaling law. Uncertainties are dominated by systematic errors, e.g. dust correction, aperture losses, which may be as high as 50\%.
\item [e] Half-light radius derived from GALFIT is not a true representation because of the complex galaxy structure.
\end{tablenotes}
\end{threeparttable}
      }}
\end{table*}
%~~~~~~~~~~~~~~~~~~~~~~~~~~~~~~~~~~~~~~~~~~~~~~~~~~~
\section{Optical Properties of the Host Galaxies of ASKAP FRBs }
Between 2018 Sept $-$ 2019 June, four FRBs were localised with ASKAP, each of which was unambiguously associated to a host galaxy.
Each burst fell within $1''$ of an $r < 22$\,mag galaxy, with an estimated chance localization of only
$\approx 0.3\%$ for a single object \citep{X+19}, or $\approx 10^{-10}$ for the combined set. Survey and targeted follow-up imaging and spectroscopy data in optical and near-IR passbands were collected to measure the redshifts and other properties of the host galaxies. 

Although public imaging survey data such as from the Dark Energy Survey \citep[DES]{DES} and Sloan Digital Sky Survey \citep[SDSS]{SDSS} were available for most host galaxies, additional follow-up imaging was conducted using the FOcal Reducer/low dispersion Spectrograph 2 \citep[FORS2]{FORS2}, X-shooter \citep{X-Shooter} mounted on the European Southern Observatory's Very Large Telescope (VLT) and Sinistro \footnote{https://lco.global/observatory/telescopes/1m/} instrument mounted on a 1~metre telescope at the Las Cumbres Observatory \citep[LCOGT]{lcogt}. Optical spectroscopy of the host galaxies was conducted using the Keck Cosmic Web Imager instrument \citep[KCWI]{KCWI} on the W. M. Keck telescope; the Gemini Multi-Object Spectrograph \citep[GMOS]{GMOS} mounted on the Gemini-South telescope; FORS2 at the VLT; and the Magellan Echellette  spectrograph \citep[MagE]{MagE} on the Magellan Baade telescope. 

We derived properties for each of the host galaxies such as stellar mass, star formation rate (SFR), internal extinction, and colors from the optical data (see Table~\ref{tab:host}). VLT imaging was reduced with {\sc esoreflex} \citep{ESOReflex} or {\sc ccdproc} \citep{ccdproc}, individual frames were co-added with {\sc montage} \citep{Montage} and photometry was performed using {\sc sextractor} \citep{SExtractor}, as described in \citet{X+19}. We corrected for Galactic extinction using the \citet{SandF} law and interpolation of extinction values using the IRSA Dust
tool\footnote{https://irsa.ipac.caltech.edu/applications/DUST/}. 
We then used the CIGALE package \citep{cigale} to perform SED fits of the data to obtain
the aforementioned galaxy properties. 
The spectra were analyzed with the pPXF package \citep{ppxf} to measure nebular line ratios and draw inferences on excitation state of the gas. \cite{Bannister+19}, \cite{X+19} and \cite{JP+19} give the full details of the procedure, and the complete dataset of measurements and derived quantities are available in the GitHub FRB repository\footnote{https://github.com/FRBs/FRB}.

Table \ref{tab:host} lists a range of measured and derived properties for the ASKAP FRBs, as well as for the only two other (known to date) localized FRBs: 121102 \citep{VLAlocalisation} and 190523 \citep{Ravi+19}. We now briefly summarize relevant aspects of each ASKAP FRB host galaxy.

\subsection{FRB~180924}
FRB~180924 was detected on 2018 September 24 at UT 16:23:12 and has a DM of $362.4\pm 0.2$ pc cm$^{-3}$. It was localized to 
a luminous, quiescent galaxy, DES~J214425.25$-$405400.81 at a redshift of $z=0.3214$ based on well-detected \oii\,,\oiii\,, H$\alpha$, H$\beta$ emission and stellar Ca~H$+$K absorption spectral lines using the KCWI and Gemini-S/GMOS spectrograph. The \oiii\ and high \nii\ /H$\alpha$ flux ratio indicate low-ionization narrow emission-line region (LINER) emission \citep{yan12}. A detailed description of the follow-up observations
and properties of the host galaxy of FRB~180924 
is given in \cite{Bannister+19}.  

Analysis of the Multi Unit Spectroscopic Explorer (MUSE) \citep{MUSE} data from the VLT obtained on 2018 Nov 5 UT shows that the ionised gas is distributed throughout the host galaxy, including
 at the position of FRB\,180924. The H$\alpha$ emission-line flux at the burst position is similar to the average H$\alpha$ emission at the same radius in the host, so the spectroscopic data show no evidence for any enhanced star formation at the position of the FRB although we are spatially limited by the seeing of the observations, with effective point spread function (PSF) of $0.8^{''}$. 

\subsection{FRB~181112}
FRB~181112 was detected on 2018 November 12 at UT 17:31:15 and has a DM of $589.0 \pm 0.3$ pc cm$^{-3}$. It is
associated with the galaxy DES~J214923.66$-$25815.28. On UT 2018 Dec 5, observations of the host of FRB~181112 with the VLT/FORS2 spectrograph established the redshift of the galaxy to be $z=0.4755$ from nebular emission lines such as \oiii\,, \nii\,, H$\alpha$ and H$\beta$ indicating ongoing star-formation. See \cite{X+19} for further details of this host galaxy, and a foreground galaxy whose halo was intersected by the FRB.

\subsection{FRB~190102}
FRB~190102 was detected on 2019 January 02 at UT 05:38:43 and has a DM of $364.5\pm 0.3$ pc cm$^{-3}$ \citep{JP+19}. A host was identified in deep FORS2 imaging performed in $g$ and $I$ band with the VLT on 2019 January 12 UT. 

On UT 2019 March 12, we observed the host galaxy with the Magellan/MagE spectrograph and obtained a redshift of $z=0.29$ from the \oii\ doublet (resolved) and other nebular emission lines, 
which was also confirmed in our VLT/FORS2 spectrograph observations on 2019 March 25, from H$\alpha$ emission. 
Its photometry and nebular line emission indicate a star-forming galaxy with a modest star-formation rate and stellar mass (Table~\ref{tab:host}).

\subsection{FRB~190608}
FRB~190608 was detected on 2019 June 08 at UT 22:48:12 and has a DM of $339.5 \pm 0.5$ pc\,cm$^{-3}$ \citep{JP+19}. It was localized to the star-forming galaxy SDSS~J221604.90$-$075355.9 at a redshift $z=0.118$, obtained from SDSS data release 9 \citep{Ahn2012}.
This is a relatively well-studied galaxy, which is known to host a Type 1 (broad-line) AGN \citep{Stern2012a}. 
Our derived SFR for this galaxy (1.24\,M$_\odot$\,yr$^{-1}$, as listed in Table \ref{tab:host}) 
agrees reasonably well with the SDSS DR12 \citep{sdss12} value of $1.7\pm0.2$\,M$_\odot$\,yr$^{-1}$. 

On UT 2019 Aug 21, we performed deep g-band imaging of this host with the X-shooter instrument at the VLT. These observations show spiral arms which were too faint to be observed in the SDSS data. Figure \ref{fig:others} shows an image of this and the other host galaxies.

\subsection{Potential biases in FRB detection }
\label{sec:detectionbiases}
There are features in FRB search algorithms that can bias their detection. 
The detectability of an FRB is a complex function of the FRB fluence, the pulse profile, and the DM. The detectability declines with increasing DM due to ``DM smearing''. Intrinsic pulse width will also reduce the signal to noise (S/N) of the burst below the triggering threshold \citep{Connor+19}.

The limits on both DM and fluence will bias against high-$z$ FRB events and we recognise that the sample studied here only reflects the $z<0.5$ FRB population. 
Also, the observed FRBs may be biased against occurring in highly turbulent environments (e.g.\ active star-forming regions) where the signal would be temporally broadened by scattering \citep{JP}.  
%~~~~~~~~~~~~~~~~~Figure 2~~~~~~~~~~~~~~~~~~~~~~~~~~~

\begin{figure*}
\centering
\includegraphics[width=1\textwidth]{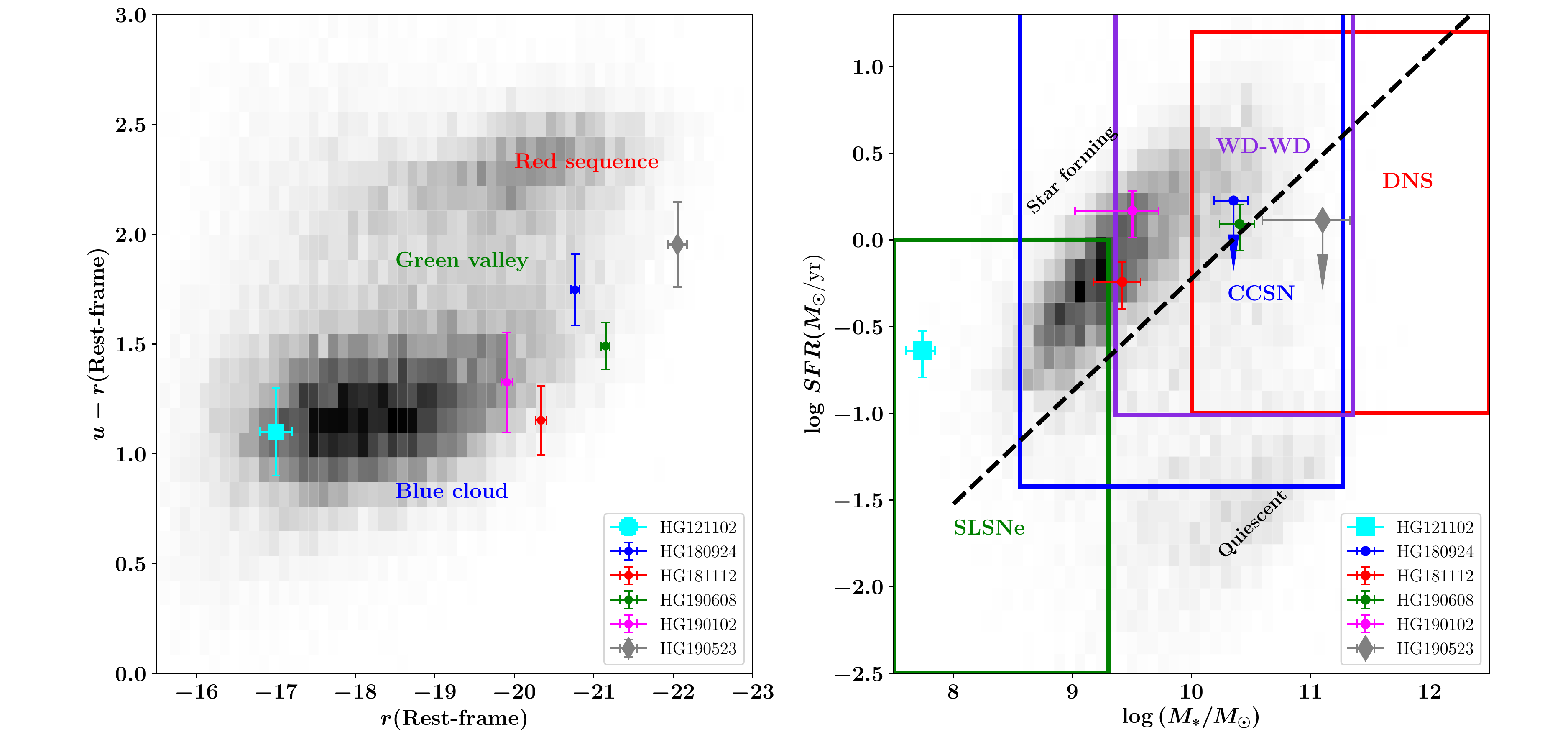}
\caption{(left) Rest-frame color-magnitude diagram of the host galaxies (HG) of the repeater (FRB\,121102), FRB~190523 and ASKAP FRBs compared to the population of $z \sim 0.3$ galaxies taken from the PRIMUS survey \citep{PRIMUS}. Note that the $u-r$ color for FRB~121102 is a notional value based on its star-forming properties. (right) Distribution of SFR vs.\ stellar mass for PRIMUS galaxies at $z \sim 0.3$ 
compared against values for the host galaxies of repeater, FRB~190523 and ASKAP FRBs. Upper limits on SFR are plotted for the hosts of FRB 180924 and FRB 190523. The dashed line separates the primary sequences of star-forming and quiescent galaxies. The green box boundary represents the 90\% region populated by the hosts of super-luminous supernovae \citep{Perley16}. The blue and violet boundaries represent the 90\% region of the galaxies hosting CCSNe and WD mergers, respectively, \citep{Kelly2012, Wolf2016}, whereas the red boundary represents the simulated galaxies which are likely to host DNS with merger rate $>10^{4}$ per galaxy per Gyr as presented in \citet{Artale19}.}
\label{fig:pop}
\end{figure*}
%~~~~~~~~~~~~~~~~~~~~~~~~~~~~~~~~~~~~~~~~~~~~~~~~~~~

\section{Radio continuum properties of the ASKAP FRB host galaxies}
We searched for persistent radio continuum emission from the host galaxies of FRB~180924, FRB~181112, FRB~190102 and FRB~190608 using 
the Australia Telescope Compact Array (ATCA, project code C3211). 
The observations were triggered within 10 days of the burst and were conducted in the 4\,cm band, with centre frequencies at 5.5~GHz and 7.5~GHz, using the ATCA's highest resolution array configurations. We observed a single epoch for FRB~180924, FRB~181112 and FRB~190608 and two epochs for FRB~190102 (See Appendix: Table \ref{tab:radio}). We combined the data from the two IF bands, and also combined the two epochs for FRB~190102, to perform a deep search of radio continuum emission at 6.5~GHz from the position of the FRB host galaxies.

We found no continuum emission from anywhere within the host galaxies of FRB~180924, FRB~181112, or FRB~190102 above a $\sim20\,\mu$Jy flux density limit (3$\sigma$). However, potentially resolved emission of $\sim 65\,\mu$Jy was observed at 5.5~GHz with a NS$-$EW resolution of $20'' \times 2''$ from the host of FRB~190608.  No radio emission was observed at 7.5~GHz above our $45\,\mu$Jy flux density limit (3\,$\sigma$) for the field of FRB~190608. The detection at 5.5~GHz was not very significant and therefore we triggered the Jansky Very Large Array (JVLA) at UT 2019 August 28 in the band spanning between 4~GHz$-$8~GHz for deeper observations (project code 19A-121). The observations were taken in the most extended ``A" configuration of the JVLA, with a resolution of $\sim$0.4 arcseconds using natural weighting. We detected resolved radio continuum emission (estimated size $\sim 2.5''$) with a peak brightness of 16 $\mu$Jy  beam$^{-1}$ (4.5~$\sigma$ significance), coincident with the nucleus of the host galaxy.  Tapering the synthesied beam to a circular Gaussian with full-width half-maximum of 0.8 arcseconds increased the detection significance to 6$\sigma$, with a peak brightness of $\sim 27~\mu$Jy beam$^{-1}$. The discrepancy between the JVLA and ATCA brightness measurements suggests that some emission is still resolved out by the JVLA observations.

The diffuse nature and probable steep spectrum of the emission suggest that this is synchrotron emission due to star formation in the host galaxy. Since we know that several of the host galaxies of our ASKAP FRBs have ongoing star formation, we can estimate the level of radio continuum emission expected from star formation alone. For this, we assume a radio-SFR relation at 1.4\,GHz of ${\rm log\,P}_{1.4} = {\rm log\,SFR}+20.95$ \citep{Sullivan2001} where $P$ is the radio luminosity. With the exception of FRB 190608, the star-formation contribution to the radio continuum emission at 6.5\,GHz is expected to be less than 1-2\,$\mu$Jy. For the closest host galaxy (FRB 190608), the expected contribution is $\sim10$\,$\mu$Jy for a SFR of $\sim 1 M_\odot$\,yr$^{-1}$ (assuming a radio spectral index of $-0.7$), which within the uncertainties, is consistent with our observations.

The 3$\sigma$ limits on radio continuum luminosity from a compact source at the locations for all four ASKAP FRBs are presented in Table \ref{tab:host}. These limits are lower than the luminosity of the FRB~121102 persistent source ($L = 1.8 \times 10^{22}$ W Hz$^{-1}$), indicating that the bursts originate from less extreme environments, if they have the same progenitors as the repeating FRB. 
From these data, we conclude that none of the four ASKAP FRB host galaxies contains a persistent compact radio source as intrinsically luminous as that seen in the host galaxy of FRB 121102. Deeper continuum searches, particularly at frequencies of a few GHz, are feasible and will provide more stringent limits or a possible detection. 
%~~~~~~~~~~~~~~~~~~~~~~ Table 2 ~~~~~~~~~~~~~~~~~~~~~~~
\begin{table*}
	\centering
	\caption{Some popular FRB progenitor models, and the stellar populations from which they arise. The information in this Table is drawn from the much larger compilation of FRB progenitor models published by \cite{FRBtheorycat}. }
	\label{tab:progenitor} 
    \begin{tabular}{lllllc}
 \hline
 & {\bf Young stellar population } & {\bf General stellar population }  & {\bf Non-stellar models} &  \\
 & (age $<10-100$\,Myr) & (all ages)  & & \\
\hline
{\bf Cataclysmic } & {\it Supernovae:} & {\it Compact-object mergers:} \\ 
(single burst) & Core-collapse SN (CCSN) & NS-NS merger (DNS) & \\ 
& SLSN/long GRB & WD-WD merger & \\
& & NS-BH merger  \\
& & BH-BH merger \\
\hline
{\bf Episodic} & {\it Magnetars: }& {\it Magnetars:} & {\it Supermassive black holes: } \\
(potential for & Young magnetar from SLSN & Magnetar from DNS merger  & AGN outburst\\
repeat bursts)  & Magnetar from CCSN & {\it White dwarfs:} & NS interaction with AGN   \\
 & {\it Pulsars: } & WD from WD-WD mergers  &  {\it Other: } \\ 
& Pulsar giant flares  & White dwarf collapse (AIC) & Superconducting cosmic strings \\
& Young SNR pulsars & NS-WD accretion \\

\hline
      \end{tabular}
\end{table*}
%~~~~~~~~~~~~~~~~~~~~~~~~~~~~~~~~~~~~~~~~~~~~~~~~~

\section{Discussion}
\subsection{Comparison of the ASKAP FRB host galaxies with the general galaxy population at $z\sim0.3$ }
We can place the ASKAP FRB hosts in context with the cosmic population by comparing their properties with a representative set of galaxies at similar redshifts. In Figure~\ref{fig:pop}, we compare the color$-$magnitude and SFR$-$stellar mass ($M_*$) distributions of the FRB host galaxies with those of the general population of galaxies at redshift $z\sim0.3$ from the PRIMUS survey \citep{PRIMUS}. 

The left panel (color-magnitude) of Figure~\ref{fig:pop} provides some information about the overall stellar population in these galaxies. The host galaxies of the ASKAP FRBs lie towards the bright end of the magnitude distribution, and mainly near the star-forming `blue cloud' \citep{Strateva2001} --- though the FRB~180924 host galaxy lies in the more sparsely-populated `green valley' region where galaxies are expected to be transitioning between star-forming and quiescent systems \citep{Martin2007b}. Although the ASKAP FRB hosts are relatively massive galaxies, none of them lie in the `red and dead` zone. 
  
The right panel of Figure~\ref{fig:pop} shows the current SFR as a function of galaxy stellar mass. The four ASKAP FRB host galaxies all have stellar masses
above a few times 10$^9$\,M$_\odot$, 
and their SFRs mainly lie on or below the star-forming main sequence for galaxies of this stellar mass --- i.e. none of these objects are `starburst' galaxies (with sSFR $>-8.7$), and only one (FRB 180924) is potentially `quiescent'. 
\subsection{Constraints on progenitor models from the host-galaxy properties of ASKAP FRBs}
Table \ref{tab:progenitor} lists some of the main FRB progenitor models, based on information from the literature as summarized by \citet{FRBtheorycat}. 

The localization of ASKAP FRBs (particularly FRB 180924 and FRB 190608) to the outskirts of their host galaxies show (i) that FRBs indeed come from galaxies, and (ii) that they are typically not coincident with the nucleus of their hosts. This information already appears to disfavor a range of models involving AGN, super-massive black holes and superconducting cosmic strings. 

In Table \ref{tab:progenitor}, we separate progenitors that are expected to arise exclusively from a young ($<10-100$\,Myr) stellar population from those that can also arise from old or intermediate-age stellar populations. 
Although the sample of localized FRBs is still small, this approach allows us to start addressing some general questions about FRB progenitors. 

If FRBs arise mainly from `young stellar population' progenitors, we would expect them to occur mainly in the kinds of host galaxies where most stars are currently forming. These are typically massive galaxies with stellar masses $10^{10} - 10^{11}$\,M$_\odot$ and SFR $>1$\,M$_\odot$\,yr$^{-1}$ \citep{Brinchmann2004}.

In contrast, if most FRBs progenitors come mainly from an old or intermediate-age stellar population (e.g. merging neutron stars), then they should mainly occur in the kinds of massive galaxies where most stars (of all ages) lie. These are typically galaxies with stellar masses above $10^{10}$\,M$_\odot$, with a wider range in current SFR. 

We now explore the expected location in the stellar mass-SFR plot (right-hand plot in Figure \ref{fig:pop}) of the host galaxies of FRBs that arise from two possible progenitor channels: (i) cataclysmic events associated with the mergers of compact objects (neutron stars, white dwarfs or stellar-mass black holes), and (ii) potentially-repeating bursts from young magnetars, which may be produced from the explosion of superluminous supernovae (SLSNe), from core-collapse supernovae (CCSNe), or potentially from the merger of two neutron stars or white dwarfs (NS-NS, see \cite{Margalit2019}; WD-WD mergers, see \cite{Levan2006}).

{\bf Host galaxies of NS, WD or BH merger events: } 
Recent simulations \citep{Artale19} predict that 70\% of simulated NS-NS mergers, 55\% of BH-NS mergers, and 53\% of BH-BH mergers occur in galaxies with stellar mass $> 10^{10} M_\odot$ and star formation rate $> 0.1 M_\odot$yr$^{-1}$ (as shown by the red-bordered region of the right plot in Figure \ref{fig:pop}). 

We find that 2/4 of ASKAP FRB hosts (and 3/5 of the localized, one-off FRB hosts in Table \ref{tab:host}) lie in a similar range of stellar mass and star formation rate. We take type Ia supernova (which are believed to arise from WD-WD mergers or a single degenerate scenario) as a proxy for all binary WD mergers. According to \citet{Wolf2016},
90\% of the host galaxies of type Ia SNe lie within the violet-bordered `WD-WD' merger region of Figure \ref{fig:pop}, in which all four of the ASKAP FRB host galaxies also lie.   \\

{\bf Host galaxies of young magnetars from SLSNe:}
High energy transients such as SLSNe and long gamma ray bursts (LGRBs) are preferentially hosted by low-mass, and high SFR dwarf galaxies \citep{Fruchter06} --- similar to the host of FRB~121102. Observations of low-redshift SLSNe \citep{Perley16} show that over 90\% of them arise in galaxies that have stellar mass $< 2\times10^{9} M_\odot$ and lie close to the star-formation main sequence. As can be seen from Figure\ \ref{fig:pop}, none of the ASKAP FRB host galaxies lie in the region where most SLSNe hosts are found. \\

{\bf Host galaxies of young magnetars from CCSNe:} Young magnetars can also be produced by the much larger population of regular core-collapse supernovae (CCSNe). According to \citet{Kelly2012}, 
90\% of the host galaxies of CCSNe, which are mostly massive with log $M_{*}/ M_{\odot}$ $> 9.5$, lie within the blue-bordered `CCSN' region in Figure \ref{fig:pop}, and all four of the ASKAP FRB host galaxies also lie in this region. As the FRB samples grow, potential biases in CCSNe selection may become important in the near future. \\

From Figure \ref{fig:pop}, we conclude that models in which all FRBs come from SLSNe/long GRB progenitors appear highly unlikely.  The host galaxies of the four localized ASKAP FRBs in Table 1 are all `normal' massive galaxies of the kind that would be expected if most FRBs come from the general stellar population. 

From host galaxy considerations alone, several FRB progenitor models still appear plausible: (i) magnetars from CCSNe; (ii) magnetars from NS-NS mergers \citep{Margalit2019} or WD-WD mergers \citep{Levan2006}; (iii) cataclysmic FRBs from NS-NS mergers, WD mergers, WD accretion mechanisms, BH mergers or a combination of these. 

%~~~~~~~~~~~~~~~~~~~ Figure 3 ~~~~~~~~~~~~~~~~~~~~~
\subsection{Energy range of the ASKAP FRBs}
\begin{figure}
\includegraphics[width=0.45\textwidth]{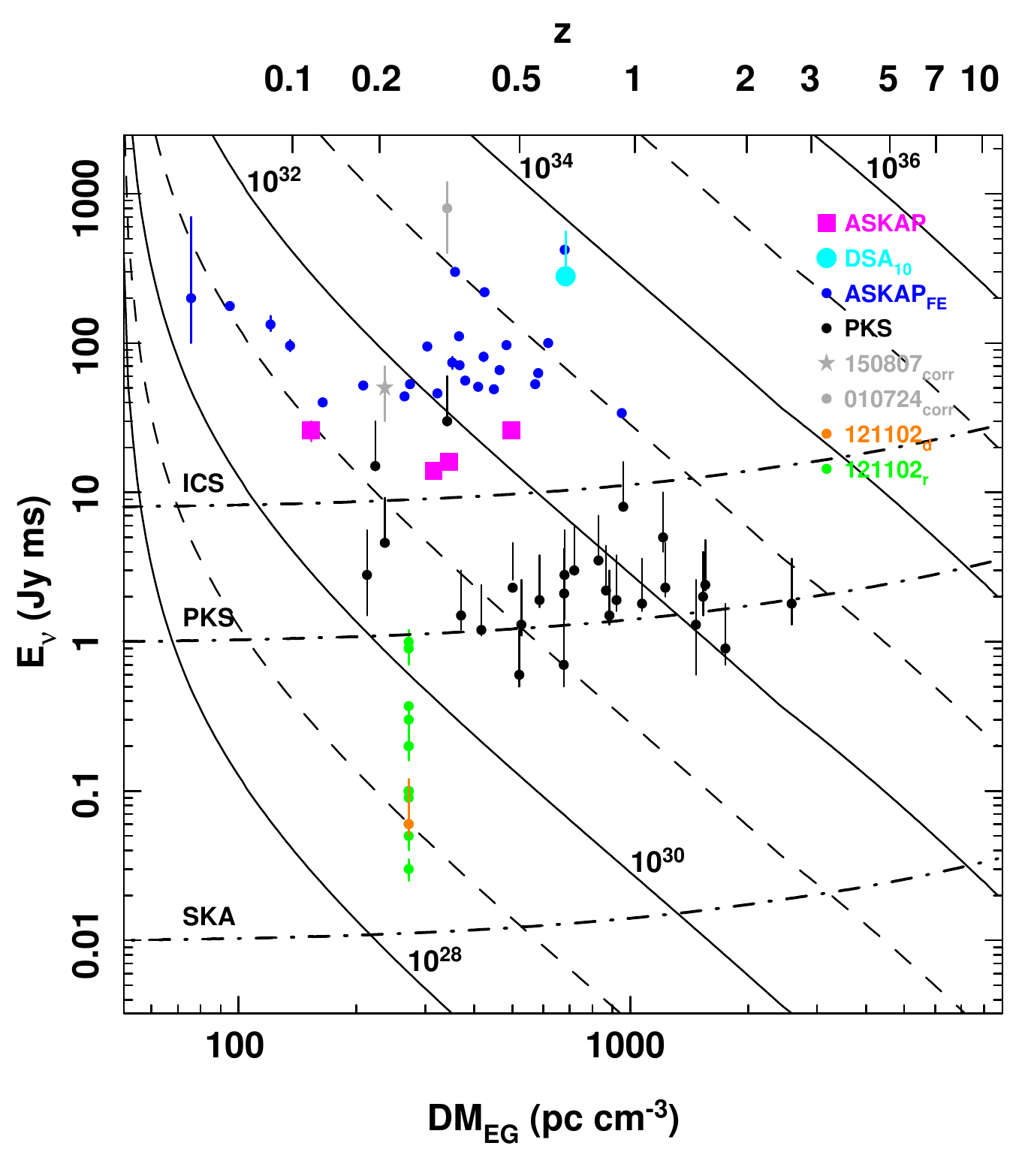}
\caption{Distribution of fluences and energies for FRBs detected in the fly's eye survey with ASKAP (blue data points), with the Parkes radio telescope (black data points), beam-corrected fluences for two Parkes FRBs (grey data points), FRB~190523 detected with DSA-10 (cyan data point) and repeating bursts detected from FRB~121102 (original burst in orange and repeating bursts in green);
adapted from Figure 2 of \citet{Shannon2018}. Magenta data points are the sample of localized ASKAP FRBs detected in incoherent searches. The black curves show contours of constant spectral energy density, in units of erg Hz$^{-1}$. The upper horizontal axis shows the redshift assuming a homogenously distributed intergalactic plasma, and a host contribution to the dispersion measure of $50(1 + z)^{-1}$ pc cm$^{-3}$ \citep{Shannon2018}. Non-localised FRBs are plotted based on their excess DM using the lower horizontal axis, while localised FRBs are plotted based on their spectroscopic redshifts and the upper horizontal axis. }
\label{figure:frb_energy}
\end{figure}
%~~~~~~~~~~~~~~~~~~~~~~~~~~~~~~~~~~~~~~~~~~~~~~~

As can be seen from Table \ref{tab:host} and Figure \ref{figure:frb_energy}, the localized ASKAP FRBs fall within the energy range ($2\times10^{30}$ to $2\times10^{34}$ erg\,Hz$^{-1}$) spanned by the ASKAP and Parkes FRBs studied by \cite{Shannon2018}. Any conclusions we can draw about the nature of the host galaxies and progenitors of these localized bursts should therefore be applicable to the general population of FRBs within this energy range, but cannot necessarily be extrapolated to any population of weaker bursts with energies below about $2\times10^{30}$ erg\,Hz$^{-1}$. 

\subsection{Is the host galaxy of FRB~121102 atypical?} 
Our analysis of the properties of ASKAP-localised FRBs and their hosts has shown that 
the repeating source FRB~121102 is anomalous. It is different in its energetics and polarisation properties as compared to the ASKAP FRB population. The energies of most of the published repeating bursts from the source of FRB 121102 
are 10$-$100 times lower than any of the ASKAP FRBs.
The RM observed for FRB~121102 is a thousand times higher than for any of the ASKAP FRBs, suggesting an unusual (or at least atypical) environment for this repeating FRB. In terms of the radio properties of host galaxies, none of the ASKAP host galaxies have a compact persistent radio source as luminous as the one in the FRB 121102 host galaxy, despite the ASKAP bursts being at least a hundred times more energetic. 

 The host galaxy for FRB~121102 is a low-mass dwarf and has an elevated SFR for its stellar mass. As \citet{Bannister+19} pointed out (their Table S8), fewer than 1\% of stars lie in galaxies as faint as this. Since the nature of the ASKAP FRB hosts implies that most FRBs are drawn from the general stellar population, the location of an FRB in such a small galaxy is (in hindsight) surprising.

Lastly, the ASKAP bursts have not been seen to repeat at the rate observed for FRB~121102 \citep{James+2019}. It seems very unlikely that FRB~121102 is drawn from the same population as the ASKAP FRBs.

\subsection{Next steps and future work}
The sample of four localized FRBs has enabled us to study the global properties of their host galaxies. A direct measurement of the ionised missing baryons along the line of sight of these FRBs is also performed in \citet{JP+19}. At least arcsecond level localization, as demonstrated by ASKAP FRBs, is required not only for confident host galaxy associations but also to study FRB environments using deep integral field unit (IFU) and imaging observations. A much larger sample of the host galaxies for FRBs is required to better characterise the host galaxy population and the distribution of FRBs within the host galaxies. 

\section{Conclusions}
The galaxy colors and star formation rates of the host galaxies of ASKAP FRBs show a diversity of properties and are not confined in a well defined locus of a particular class. They exhibit lower star formation relative to their high stellar mass, very different to the host of FRB~121102, which is a starburst dwarf galaxy. Additionally, no persistent co-located compact radio sources were detected at the level seen in the host of FRB~121102.

The arcsecond localization of most ASKAP FRBs, and sub-arcsecond localization of FRB~180924, confirms that they occur in the outer regions of their hosts, ruling out the models involving AGNs and superconducting cosmic strings for all FRBs. 

The global properties of the host galaxies of ASKAP FRBs suggest that the broader population of FRBs can arise from both young and (moderately) old progenitors. The considerations of the host galaxy properties
make SLSNe less likely to be their progenitors, while WD-WD and NS-NS mergers, accretion-induced WD collapse and regular CCSNe seem to be plausible mechanisms for at least a subset of the FRB population.     

\section*{Acknowledgements}
\begin{figure}
\begin{tabular}{cc}
\includegraphics[width=7.5cm]{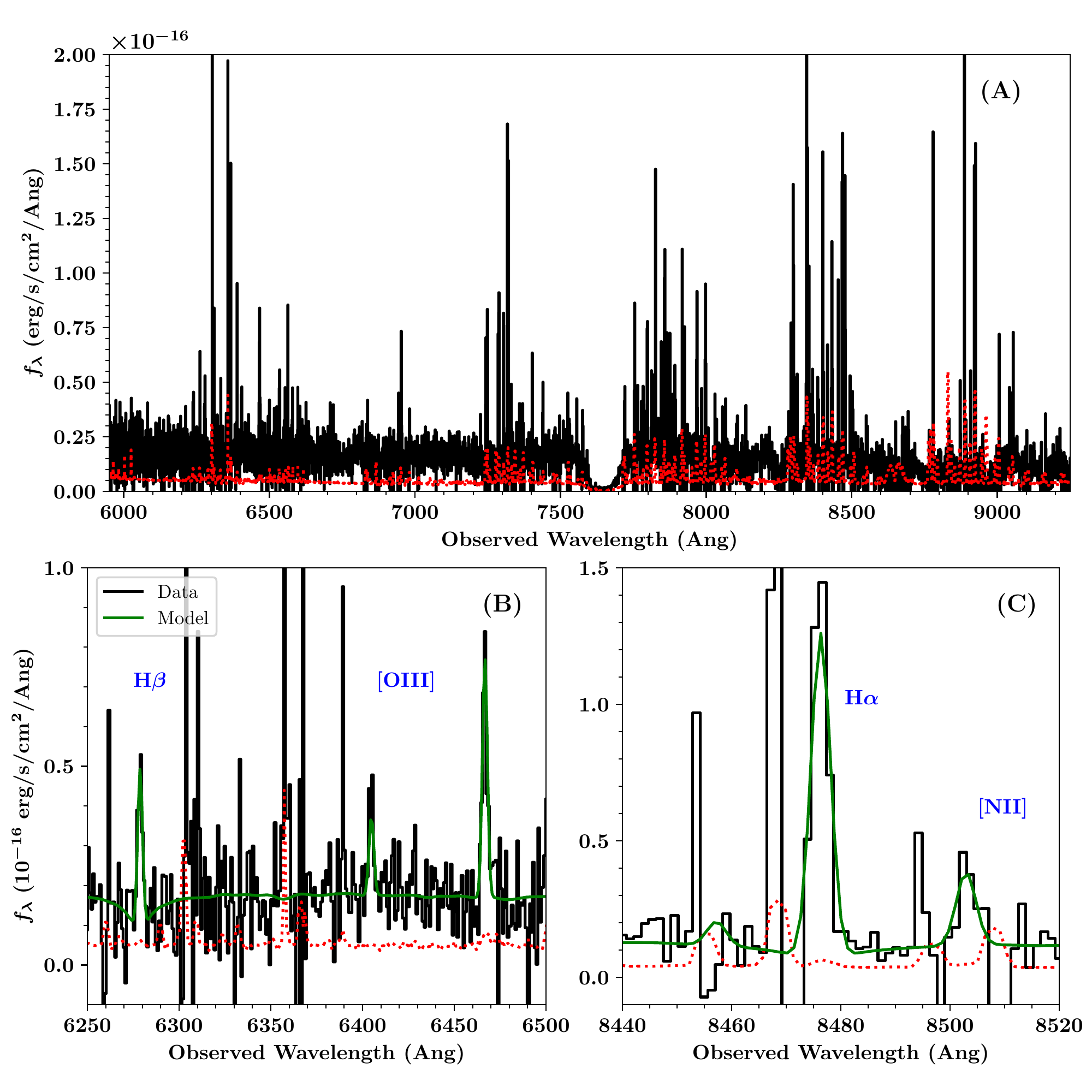} &
\includegraphics[width=9.5cm]{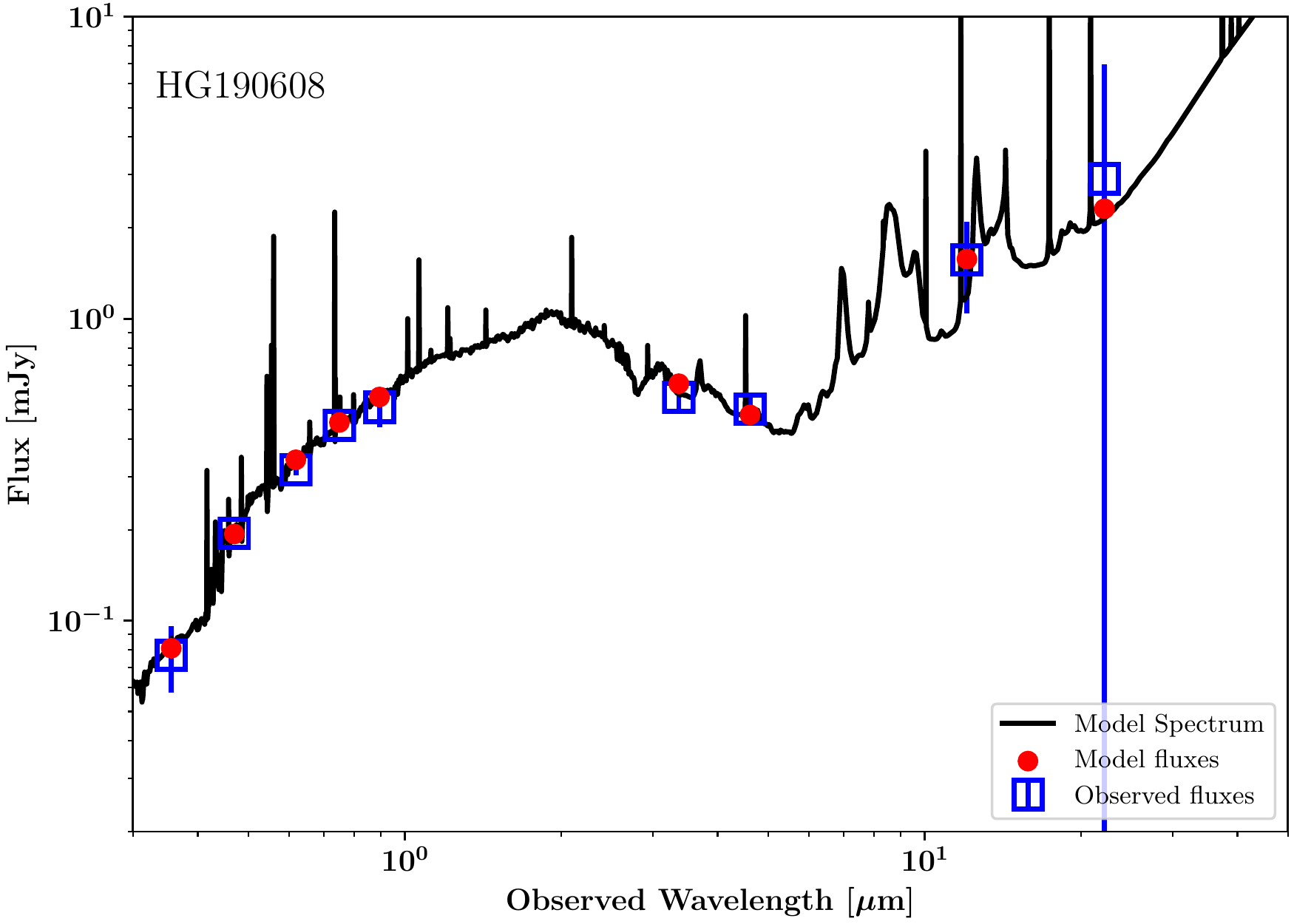} \\

\end{tabular}
\caption{Left panel: Spectrum of the host galaxy of FRB~190102. 
Right panel: SED for the host of FRB~190608. }
\label{fig:seds}
\end{figure}

Based on observations collected at the European Southern Observatory under ESO programmes 0102.A-0450(A) and 0103.A-0101(B). K.W.B., J.P.M, and R.M.S. acknowledge Australian Research Council (ARC) grant DP180100857.
 A.T.D. is the recipient of an ARC Future Fellowship (FT150100415).
 S.O. and R.M.S. acknowledge ARC grant FL150100148. R.M.S. also acknowledges support through ARC grant CE170100004. N.T. and F.C.G. acknowledges
support from PUCV/VRIEA project 039.395/2019. SL was funded by projects UCh/VID-ENL18/18 and FONDECYT 1191232.
J.X.P. and S.S. are supported by NSF AST-1911140.
The Australian Square Kilometre Array Pathfinder and Australia Telescope Compact Array are part of the Australia Telescope National Facility which is managed by CSIRO. 
Operation of ASKAP is funded by the Australian Government with support from the National Collaborative Research Infrastructure Strategy. ASKAP uses the resources of the Pawsey Supercomputing Centre. Establishment of ASKAP, the Murchison Radio-astronomy Observatory and the Pawsey Supercomputing Centre are initiatives of the Australian Government, with support from the Government of Western Australia and the Science and Industry Endowment Fund. 
We acknowledge the Wajarri Yamatji as the traditional owners of the Murchison Radio-astronomy Observatory site. The National Radio Astronomy Observatory is a facility of the National Science Foundation operated under cooperative agreement by Associated Universities, Inc. 
Spectra were obtained at the W. M. Keck Observatory, which is operated as a scientific partnership among Caltech, the University of California, and the National Aeronautics and Space Administration (NASA). The Keck Observatory was made possible by the generous financial support of the W. M. Keck Foundation. The authors recognize and acknowledge the very significant cultural role and reverence that the summit of Mauna Kea has always had within the indigenous Hawaiian community. We are most fortunate to have the opportunity to conduct observations from this mountain. This paper includes data gathered with the $6.5$ meter Magellan Telescopes located at Las Campanas Observatory, Chile, as part of program CN2019A-36. This work makes use of observations from the LCOGT network obtained as part of programs CN2019A-39/CLN2019A-002 and CN2019B-93/CLN2019B-001. The Gemini-S/GMOS observations were carried out as part of program GS-2018B-Q-133, obtained at the Gemini Observatory, which is operated by the Association of Universities for Research in Astronomy, Inc., under a cooperative agreement with the NSF on behalf of the Gemini partnership: the National Science Foundation (United States), National Research Council (Canada), CONICYT (Chile), Ministerio de Ciencia, Tecnolog\'{i}a e Innovaci\'{o}n Productiva (Argentina), Minist\'{e}rio da Ci\^{e}ncia, Tecnologia e Inova\c{c}\~{a}o (Brazil), and Korea Astronomy and Space Science Institute (Republic of Korea).
\bibliography{references.bib}
\appendix
\textbf{SED fitting using CIGALE}: To estimate the stellar mass, star formation rates and restframe colors of the host galaxies for FRB~180924, FRB~181112 and FRB~190608, we used CIGALE \citep{cigale}, a python-based SED fitting software. The SED for the host of FRB~180924 and FRB~181112 are presented in \citet{Bannister+19} and \citet{X+19} and an SED for the host of FRB~190608 is presented in Fig.\ref{fig:seds}. CIGALE is fed photometric measurements and redshifts and it computes optimal galaxy properties for assumed models of star formation history, stellar population, AGN emission, dust attenuation, and dust emission using a Bayesian framework. We mostly rely on optical photometry but include WISE measurements where available (Table \ref{tab:photom}). Due to degeneracies in the model parameters, the estimated properties are somewhat poorly constrained i.e. have errors $\gtrsim0.3$ dex. We use the following modules in CIGALE:
\begin{itemize}
    \item \textbf{Star formation history (SFH)}: \texttt{sfhdelayed}. It is a delayed-exponential SFH with an initial linear increase and subsequent exponential decay. We do not include a burst population and limit the maximum allowed age of the main population to the cosmological time at the galaxy's redshift. We fix the initial SFR to 0.1 $M_\odot yr^{-1}$. We allow the e-folding time of the SFH to vary but try to limit it such that the CIGALE estimate for the final SFR is consistent with our spectroscopic estimate. 
    \item \textbf{Stellar population model}: \texttt{bc03}. A stellar population model from \citet{2003MNRAS.344.1000B}. We use a Chabrier initial mass function and the separation between young and old stars is set to $10^7$ yr.
    \item \textbf{Dust attenuation}: \texttt{dustatt\_calzleit}. Dust attenuation model from \citet{Calzetti_2000} and \citet{2002ApJS..140..303L} for the optical and UV wavelengths respectively. We fix the UV bump centroid to 217.5 nm, its amplitude to 1.3 and set its FWHM to 35.6 nm. We also set the index of the power-law $(\delta)$ modifying the attenuation curve to -0.38.
    \item \textbf{Dust emission}: \texttt{dale2014}. IR, submillimeter and radio emission templates from \citet{2014AAS...22345301D}. We allow both the AGN fraction and the template parameter $\alpha$ to vary in our fits. This emission is poorly constrained in the absence of WISE measurements.
\end{itemize}
To estimate the galaxy properties for the host of FRB~190102, the spectra were analyzed with the pPXF package \citep{ppxf}. The spectrum of the FRB~190102 is also presented in Fig.\ref{fig:seds}. 

\begin{figure}[!ht]
\centering
\includegraphics[width=0.45\textwidth]{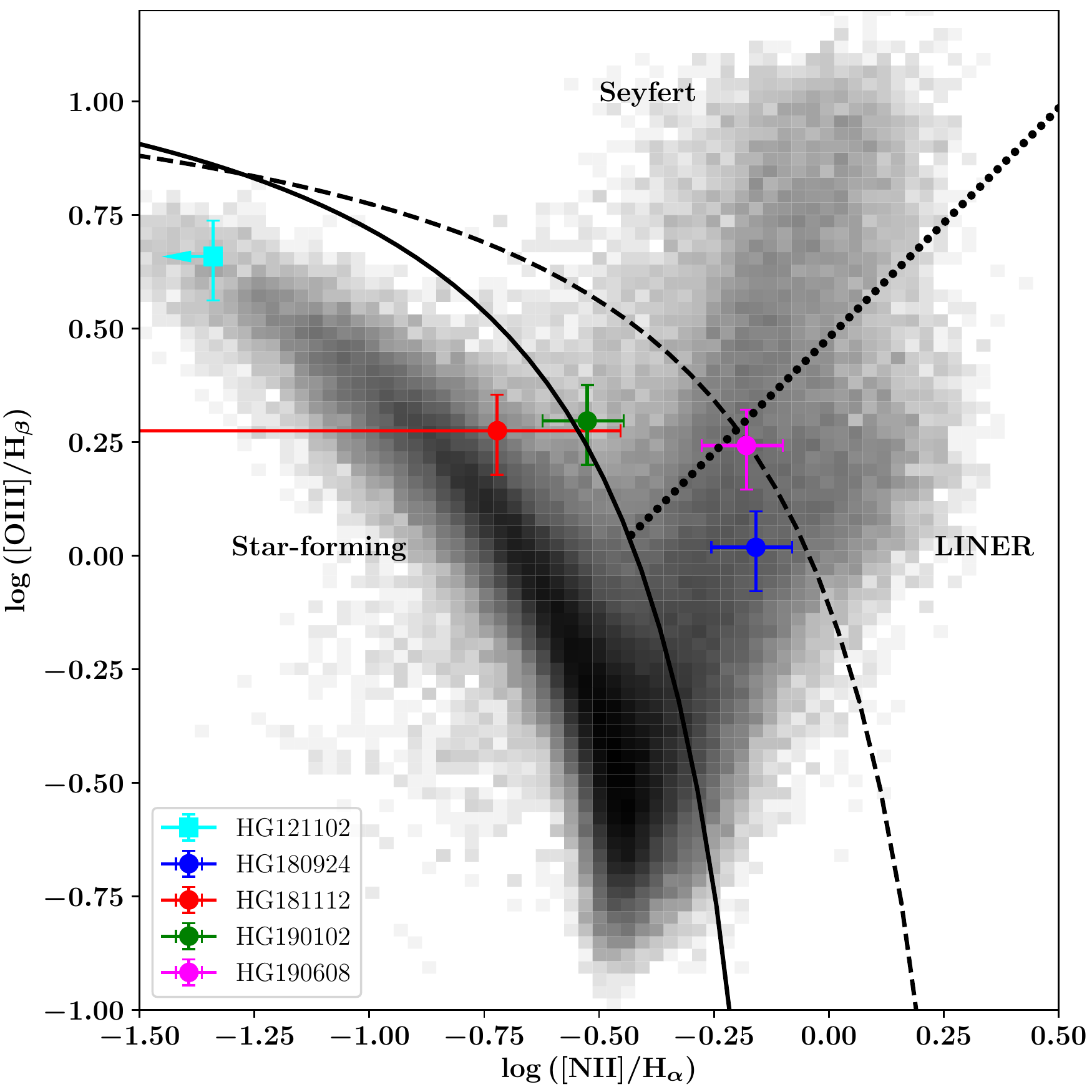}
\caption{
\citet{BPT81} diagnostic plot showing emission line ratios for FRB host galaxies (abbreviated to HG). The background data points show the distribution of $\sim 75,000$ nearby ($0.02 < z < 0.4$) emission-line galaxies from the Sloan Digital Sky survey, restricted to have S/N $> 5$. Black lines separate the star-forming galaxies (solid) from sources dominated by hard spectra (dashed), and the dotted line separates sources designated as AGNs into either Seyfert or LINER galaxies. 
}
\label{figure:fig}
\end{figure}

Figure~\ref{figure:fig} shows a BPT diagram \citep{BPT81} of the galaxy distribution for two nebular line ratios, which can be used to determine the dominant source of ionization for the interstellar medium. Only one of the four ASKAP FRB host galaxies (HG~181112) lies close to the region populated by star-forming galaxies, while HG~190102 lies in the `composite' region close to the SF/AGN boundary. The FRB~180924 host lies in the LINER region, where gas may be ionized either by a low-excitation AGN or by a population of post-AGB stars \citep{yan12}. Interestingly, the host of FRB~190608 lies at the boundary of Seyfert and LINER. It is a known Type 1 AGN (i.e. Seyfert) as noted before. Thus the four ASKAP host galaxies show a diversity of ionization properties, rather than being drawn from a population of purely star-forming systems. For further comparison, we include measurements from the host of FRB~121102, which lies in the star-forming population. 

%~~~~~~~~~~~~~~~~ SEDs ~~~~~~~~~~~~~~~~~~~
%~~~~~~~~~~~~~~~~~~~~~~~~~~~~~~~~~~~~~~~~~~~~~~~~~~~

The details of the radio and optical follow-up observations for the hosts of ASKAP localised FRBs are listed in Table \ref{tab:radio} and Table \ref{tab:optical} respectively.
%~~~~~~~~~ Used version ~~~~~~~~~~~~~~~~
\begin{table}
\caption{Photometric details for ASKAP localised host galaxies}
 	\resizebox{19cm}{!}{
	\subfloat[Photometry for HG180924 and HG181112 respectively]{
	\hspace*{-5.0cm}
		%\caption{Photometry for HG180924}
		\begin{tabular}{ccccc|cccc|cc|ccc}
		\hline
         \multicolumn{5}{c}{DES} & \multicolumn{4}{c}{WISE} & \multicolumn{2}{c}{VLT} & \multicolumn{3}{c}{LCOGT}\\
        \hline
         g & r & i & z & Y & W1 & W2 & W3 & W4 & g & I  & g & r & i \\
	 21.62(3) & 20.54(2) & 20.14(2) & 19.85(2) & 19.81(6) & 16.85(10) & 16.06(18) & 11.69(-) & 8.50(-) & 21.38(4) & 20.10(2)  & 21.59(12) & 20.46(8) & 20.20(11) \\
	 22.71(9) & 21.73(5) & 21.49(6) & 21.45(11)  & 21.07(17) & - & - & - &-  & 22.57(4) & 21.51(4) & 22.37(33) & 21.61(22) & 21.29(26)\\
		\end{tabular}
		%\label{tab:hg180924}
		}}
	
	   %\subfloat[Photometry for HG181112]{
		%\centering
		%\caption{Photometry for HG181112}
		%\begin{tabular}{ccccc|cc|ccc}
	    %\hline
         %\multicolumn{5}{c}{DES} & %\multicolumn{2}{c}{VLT} & %\multicolumn{3}{c}{LCOGT}\\
        %\hline
        %g & r & i & z & Y  & g & I & g$^{'}$ & r$^{'}$ & i$^{'}$ \\
		 %22.71(9) & 21.73(5) & 21.49(6) & 21.45(11) & 21.07(17)  & 22.57(4) & 21.51(4) & -& - & -  \\
		%\end{tabular}
		%%\label{tab:hg181112}
	%}
	
	\subfloat[Photometry for HG190102]{
		\centering
		\hspace*{-3.3cm}
		%\caption{Photometry for HG190102}
		\resizebox{11cm}{!}{
		\begin{tabular}{cccc|ccc}
		\hline
        \multicolumn{4}{c}{VLT} & \multicolumn{3}{c}{LCOGT}\\
        \hline
        u & g & I & z & g& r & i \\
	23.7(2) & 22.6(1) & 21.10(5) & 20.8(2) & $> 22.13$ & 22.06(51)  &22.02(73)\\
		\end{tabular}
		}
		%\label{tab:hg190102}
	}
	\subfloat[Photometry for HG190608]{
		%\centering
		\hspace*{-2cm}
		%\caption{Photometry for HG190608}
		\resizebox{12cm}{!}{
		\begin{tabular}{ccccc|ccc}
	    \hline
         \multicolumn{5}{c}{SDSS} & \multicolumn{3}{c}{LCOGT}\\
        \hline
        u & g & r & i & z & g & r & i\\
		  19.19(9) & 18.18(2) & 17.65(1) & 17.28(2) & 17.13(5) & 18.28(2) & 17.38(1) & 17.55(2)  \\
		\end{tabular}
		%\label{tab:hg190608}
	}}
	
	\label{tab:photom}
\end{table}

\begin{table*}
 \caption{Radio follow-up observations of the host galaxies of ASKAP localised FRBs.}
    \label{tab:radio}
    \begin{tabular}{cccccccc}
        \hline
    FRB &     Telescope        & Band    & $T_{\mathrm{start}}$  & $T_{\mathrm{obs}}$    & $T_\mathrm{start} - T_\mathrm{FRB}$   & Resolution & Rms noise\\ 
                    &               &                           (GHz)                   & (UTC)                 & (s)                   & (dd:hh:mm:ss)  & (NS$\times$EW) & ($\mu$Jy/beam)                       \\ \hline

    180924 &     ATCA-6A                 & $4.5-8.5$         & 2018-10-04-12:09:44   & 15300             & 009:19:46:33
     &    $8^{''}\times2^{''}$ & $7$ \\
     181112&     ATCA-6B                     & $4.5-8.5$    & 2018-11-17-02:47:04   & 34200             & 004:09:15:50
       & $3^{''}\times2^{''}$ & $5$\\
   190102 &      ATCA-1.5D                  & $4.5-8.5$      & 2019-01-10-12:21:14   & 21600             & 008:06:42:32 &
   $5^{''}\times2^{''}$ & $10$ \\
         
   &      ATCA-1.5D                  & $4.5-8.5$        & 2019-01-14-05:43:24   & 22320             & 012:00:04:42 &
   $4^{''}\times2^{''}$ &  $10$ \\
         
         190608 & ATCA-6A                  & $4.5-8.5$  & 2019-06-16:13:43:35 & 9000 & 017:14:55:22& $20^{''}\times2^{''}$ & $15$ \\
         
         & VLA-A                  & $4.0-8.0$  & 2019-08-25:05:40:54 & 3036 & 077:06:52:41& $1^{''}\times1^{''}$ & $3.5$
         \\
      \end{tabular}
\end{table*}   

%~~~~~~~~~~~~~~ Optical Observation ~~~~~~~~~~~~~~~~~~~
    \begin{table*}
    \caption{Optical follow-up observations of the host galaxy of FRB 180924; $T_\mathrm{FRB}$ = 2018-09-24-16:23:12, FRB 181112; $T_\mathrm{FRB}$ = 2018-11-12-17:31:15, FRB 190102; $T_\mathrm{FRB}$ = 2019-01-02-05:38:43 and FRB 190608; $T_\mathrm{FRB}$ = 2019-06-08-22:48:13.
    $T_{\mathrm{obs}}$ is the length of observation, with multiple exposures signified by a multiplication.}
    \label{tab:optical}
    \resizebox{\textwidth}{!}{
    \hspace*{-3.0cm}
    \begin{tabular}{ccccccccc}
    
        \hline
        Telescope  & Instrument    & Observation Mode  & Band    & Effective Wavelength & $T_{\mathrm{start}}$  & $T_{\mathrm{obs}}$    & $T_\mathrm{start} - T_\mathrm{FRB}$   \\ 
                    &               &                   &       & ($10^{-9}$m)                   & (UTC)                 & (s)                   & (dd:hh:mm:ss)                         \\ \hline 
                    \\
               %& & &      \textbf{Host of FRB~180924}    \\
             &   &  &  &\textbf{Host of FRB~180924} &  &   \\
               
       Keck & KCWI & IFU & Optical & $350-550$ &  2018-10-04 & $4 \times 600 $ &  10 \\
       Gemini-S   & GMOS      & Spectroscopy  & Optical   & $470-930$ &2018-10-05-02:15:48 &     $4 \times 700$                              &  010:09:52:36
         \\
         
       VLT        & MUSE     & IFU       & Optical        & $475-930$  & 2018-11-05   & $4 \times 628 $     & 41
         \\ 
         
        VLT        & FORS2     & Imaging       & g         & $470$      & 2018-11-09-01:02:49   & $5 \times 500 $   & 045:08:39:36
         \\
        VLT        & FORS2     & Imaging       & I         & $768$      & 2018-11-09-01:48:09   & $5 \times 90 $    & 045:09:24:56
         \\
          LCOGT-1m      & Sinistro  & Imaging       & i        & $754.5 $    &  2019-05-31-15:38:14           & $6 \times 60$    & 248:23:15:02
         \\
         LCOGT-1m      & Sinistro  & Imaging       & r        & $621.5 $    &  2019-05-31-15:48:36           & $10 \times 60$    & 248:23:25:24
         \\
         LCOGT-1m      & Sinistro  & Imaging       & g        & $477 $      &  2019-05-31-16:03:20  & $10 \times 60$    & 248:23:40:08
         \\
         VLT  & X-Shooter          & Imaging       & g      & $477$  & 2019-08-21-04:39:36   & $9 \times 300$    & 330:10:52:38 \\
        VLT  & X-Shooter          & Imaging       & I       & $806$  & 2019-08-21-05:23:44   & $9 \times 120$    & 330:11:39:36 \\
        VLT        & FORS2     & Imaging       & g         & $470$      & 2019-08-23-04:21:33   & $5 \times 500 $   & 332:11:58:21
         \\
        VLT        & FORS2     & Imaging       & I         & $768$      & 2018-08-23-05:29:19   & $5 \times 90 $    & 332:13:06:07
         \\
 
         \\
         &   &  &  &\textbf{Host of FRB~181112} &  &   \\
              
         VLT        & FORS2     & Imaging       & g         & $470$  & 2018-12-03-01:34:13   & $5 \times 500 $   & 020:08:02:59
         \\
         VLT        & FORS2     & Imaging       & I         & $768$  & 2018-12-03-02:10:12   & $5 \times 90 $    & 020:08:38:57
         \\
         LCOGT-1m      & Sinistro  & Imaging       & i        & $754.5$&  2019-05-31-19:10:42           & $10 \times 60$    & 200:01:39:27
         \\
         LCOGT-1m      & Sinistro  & Imaging       & r        & $621.5$&  2019-05-31-19:25:25           & $10 \times 60$    & 200:01:54:10
         \\
         LCOGT-1m      & Sinistro  & Imaging       & g        & $477$  &  2019-05-31-19:40:09          & $10 \times 60$    & 200:02:08:54
         \\
         
         VLT  & X-Shooter          & Imaging       & g      & $477$  & 2019-08-21-02:02:56   & $9 \times 300$    & 281:08:31:41
         \\
         VLT  & X-Shooter          & Imaging       & I       & $806$  & 2019-08-21-02:50:01   & $9 \times 120$    & 281:09:18:46
         \\
         VLT        & FORS2     & Imaging       & g         & $470$      & 2019-08-23-04:08:18   & $5 \times 500 $   & 283:10:37:03
         \\
        VLT        & FORS2     & Imaging       & I         & $768$      & 2019-08-23-04:42:16   & $5 \times 90 $    & 283:11:11:01
         \\
         
         \\
         &   &  &  &\textbf{Host of FRB~190102} &  &   \\
         VLT        & FORS2     & Imaging       & I         & $768 $  & 2019-01-12-01:14:23   & $5 \times 90 $    & 009:19:35:41
         \\
         VLT        & FORS2     & Imaging       & g         & $470$  & 2019-01-12-01:25:48   & $3 \times 500 $   & 009:19:47:06
         \\
         Magellan   & MagE      & Spectroscopy  & Optical   & $320-900$&  2019-03-12-08:36:03   & $1 \times 3700$ & 069:02:57:20
         \\
          LCOGT-1m      & Sinistro  & Imaging       & i        & $754.5 $& 2019-05-31-16:39:09            & $11\times 60$    & 149:11:00:26
         \\
         LCOGT-1m      & Sinistro  & Imaging       & g        & $477  $  & 2019-05-31-17:16:01          & $10 \times 60$    & 149:11:37:18
         \\
         LCOGT-1m      & Sinistro  & Imaging       & r        & $621.5 $& 2019-05-31-17:30:44            & $10 \times 60$    & 149:11:52:01
         \\

         VLT        & FORS2     & Imaging       & g         & $470$  & 2019-06-02-05:43:05   & $4 \times 500 $   & 151:00:04:22
         \\ 
         
         VLT        & FORS2     & Imaging       & u         & $361$  & 2019-06-17-09:19:36   & $5 \times 560$    & 166:03:40:54
         \\
         VLT        & FORS2     & Imaging       & z         & $910$  & 2019-06-17-10:09:45   & $5 \times 30$     & 166:04:31:03
         \\
         VLT  & X-Shooter          & Imaging       & g      & $477$  & 2019-08-21-05:51:53  & $9 \times 300$    & 231:00:13:10
         \\
         VLT  & X-Shooter          & Imaging       & I       & $806$  & 2019-08-21-06:39:02   & $9 \times 120$    & 231:01:00:19
         \\
         VLT        & FORS2     & Imaging       & I         & $768 $  & 2019-08-23-03:55:01   & $5 \times 90 $    & 232:22:16:18
         \\
         VLT        & FORS2     & Imaging       & g         & $470$  & 2019-08-23-07:19:27   & $5 \times 500 $   & 233:01:40:44
         \\
         VLT        & FORS2     & Imaging       & u         & $361$  & 2019-11-20-01:25:38   & $5 \times 560$    & 321:19:36:55

        \\
         &   &  &  &\textbf{Host of FRB~190608} &  &   \\
          LCOGT-1m      & Sinistro  & Imaging       & i        & $754.5 $&  2019-06-09-03:05:10         & $10\times 60$    & 00:04:16:57\\
         LCOGT-1m      & Sinistro  & Imaging       & g       & $477  $  & 2019-06-09-03:19:52          & $10 \times 60$    & 00:04:31:39  \\
         LCOGT-1m      & Sinistro  & Imaging       & r       & $621.5  $  & 2019-06-09-03:34:34          & $10 \times 60$    & 00:04:46:21 \\
        VLT  & X-Shooter          & Imaging       & g         & $477$  & 2019-08-21-04:39:36   & $8 \times 300$    & 073:05:51:23 \\
        VLT  & X-Shooter          & Imaging       & I         & $806$  & 2019-08-21-05:23:44   & $8 \times 120$    & 073:06:35:31 \\
         %\hline
         
    \end{tabular}}
\end{table*}

\end{document}